\documentclass[a4paper, amsfonts, amssymb, amsmath, reprint, showkeys, nofootinbib, twoside]{revtex4-1}
\usepackage[english]{babel}
\usepackage[utf8]{inputenc}
\usepackage{booktabs}
\usepackage{svg}
\usepackage[table,xcdraw]{xcolor}
\usepackage[colorinlistoftodos, color=green!40, prependcaption]{todonotes}
\usepackage[pdftex, pdftitle={Article}, pdfauthor={Author}]{hyperref} 
\usepackage{adjustbox}
\usepackage{todonotes}
\usepackage{float}
\usepackage{rotating}
\usepackage{comment}
\usepackage{graphicx}
\usepackage{setspace}
\usepackage{mathtools} 
\usepackage[defaultlines=3,all]{nowidow}

\begin{document}

\title{\Large The Geoeconomics of Venture Capital\\
\large An Economic Complexity Approach to Emerging Technological Sovereignty}


\author{Benjamin Leroy}\affiliation{Agoranov, Paris, France} 

\author{Davi Marim}\affiliation{Télécom Paris, Paris, France} 

\author{El Ghali Benjelloun}\affiliation{University Dauphine-PSL, Paris, France} 

\author{Arthur Rozan Debeaurain}\affiliation{Agoranov, Paris, France\relax}
%

\author{Jean-Michel Dalle}\email[Correspondence: ]{jmd@agoranov.com}\affiliation{Agoranov, Paris, France}\affiliation{Sorbonne Université, Paris, France\\}

\date{April 11, 2026}

\begin{abstract}
We explore a quantitative approach to emerging technological sovereignty and geoeconomic power by assessing the relative positioning of countries with economic complexity methods applied to the structure of national venture-capital (VC) portfolios and their associated Revealed Venture Advantage (RVA) metrics. Using Crunchbase firm- and deal-level data, we map venture-backed startups to 18 emerging technology domains via a probabilistic multi-label large-language-model classifier, and construct an RVA-based country–technology specialization matrix for the 17 countries with the highest aggregate VC funding. From this matrix, we derive two eigenvector-based measures: a Geoeconomic Complexity Index (GCI) that ranks countries by the composition of their venture specializations, and an Emerging Technology Geoeconomic Complexity Index (ETGCI) that ranks domains by the extent to which specialization is concentrated among high-GCI countries. Empirically, Cloud Computing, Cybersecurity Tools, and Medtech exhibit the highest ETGCI values, reflecting concentration of specialization in a small set of leading countries. The United States and Israel consistently occupy a marked “high-diversity/low-ubiquity” position and lead the GCI ranking, followed by China, France, Japan and Germany; both country and domain rankings are stable from 2021–2024. Finally, relatedness-based simulations identify when it exists, for each country, the Simplest Single Sovereignty Enhancing Technology (SSSET), i.e., the most feasible single new technological direction associated with the largest expected improvement in relative geoeconomic positioning.
\end{abstract}
\maketitle

\section{Introduction}
\label{intro}
Countries are now engaged, \emph{volens nolens}, in a sovereignty game where the nature of the emerging technological domains they control  -- be it quantum computing, fusion energy, cell therapies, space technologies, artificial intelligence (AI) or robotics -- considerably determines their geoeconomic power. In this sovereignty game, a substantial share of the capabilities that countries develop is produced by the numerous venture-backed startups active in each of these emerging technological domains, namely, by the extent to which these start-up firms invest in emerging technological domains thanks to the venture capital (VC) funding they attract. As a consequence, the geoeconomic power of a country is significantly influenced by the allocation of venture capital flows across various emerging technological directions, as they crucially influence the formation and incubation \cite{agarwal} of new industries relevant to its strategic autonomy.

To put it differently, in the current "technological sovereignty game", the cards in the hands of countries significantly depend upon venture capital flows and upon their allocation: it is not just the number of these cards that matters but also their relative power -- whether they are trumps, or not. Although it might be crucial for a country to control a technology because of the specific nature of its applications -- e.g., artificial intelligence, or quantum computing because of its expected potential to crack existing cryptographic methods, etc. --, the \emph{geoeconomic power} of an emerging technology can be strategically limited if it is widespread among many countries and thus easier to procure from partners and allies compared to another technology that only a couple of geoeconomically powerful countries would strategically control. In other words, the technological sovereignty game is actually a /emph{relative} geoeconomic contest where only the latter technological directions end up being real trumps, and it is therefore a form of na\"ivet\'e to infer geoeconomic power from applications alone and to neglect the relative advantage that can be gained in the global emerging sovereignty game.

However, despite the salience of these issues in the 21st century, existing assessments have for now remained predominantly qualitative and based on anecdotal evidence, therefore severely limiting strategic analyses across either countries or technological domains. To address this gap, we develop here a framework grounded in economic complexity theory \cite{ecomplx}: following seminal applications that have related national capabilities to export baskets \cite{eci}, we study here venture-investment portfolios through similar lenses, in order to assess the strategic positioning of countries with respect to emerging technological domains. Using global VC data, we compute Revealed Venture Advantage metrics, analyze the country–technology bipartite matrix and network and derive measures of specialization, relatedness, and geoeconomic complexity. We notably construct two indices: (i) an \textit{Emerging Technology Geoeconomic Complexity Index} (ETGCI) that ranks technological domains by their contribution to the geoeconomic power of countries; and (ii) a \textit{Geoeconomic Complexity Index} (GCI) that ranks countries according to the geoeconomic power of the directions of their venture specializations across those technological domains. Taken together, these metrics enable a more comparative, evidence-based evaluation of which technological fields contribute most to geoeconomic power, and of which countries have most strategically developed sovereign portfolios, i.e., portfolios whose technological directions correspond to the most geoeconomically powerful emerging technological domains.

The remainder of the article is organized as follows. Section~\ref{sovereign} briefly discusses technological sovereignty and geoeconomic power, and the role of venture-backed startups in these respects, motivating the need for a quantitative assessment. Section~\ref{complexity} introduces the economic-complexity approach and adapts it to VC investments in emerging technological domains. Section~\ref{methods} describes the startup and VC dataset, defines the country–technology specialization matrix, and introduces the associated metrics and geoeconomic indices. Section~\ref{results} presents results on specialization, diversity and ubiquity, and the GCI and ETGCI indices across countries and technologies. Section~\ref{limitations} outlines current limitations and discusses improvements building on the first step presented in this paper. Section~\ref{conclusion} concludes.

\section{Emerging Technological Sovereignty and Relative Geoeconomic Power}
\label{sovereign}

Over the past decade, technological sovereignty has attracted growing scholarly and policy attention, driven by concerns over how technological capabilities are distributed internationally across emerging domains such as artificial intelligence, biotechnology, quantum computing, and fusion energy. Early debates were largely framed around digital sovereignty, catalyzed by the market power of major U.S. digital corporations (often grouped under labels such as GAFAM or “Big Tech”) and featured prominently in discussions in Europe \cite{digital_sov} and China \cite{new_sov}. More recently, the concept has expanded beyond the digital sphere to encompass a broader set of emerging technological domains.

In this paper, we adopt a conventional understanding of technological sovereignty as the ability of a polity to formulate and implement economic and technology policies while limiting vulnerability to external dependencies \cite{tech_sov, def_tech_sov}. In that sense, it can be seen as a necessary condition for strategic autonomy, even if its interpretation and operationalization vary across geopolitical, economic, and institutional contexts. In Europe, for instance, the emphasis on strategic autonomy has intensified in response to perceived dependencies and the need to secure critical infrastructures and innovation capabilities \cite{eu_mandate, draghi}. In the United States, policy has increasingly taken the form of selective decoupling aimed at reducing technological dependence on China. The United Kingdom has sought to limit exposure to critical import dependencies through initiatives such as “Project Defence,” while China has emphasized self-reliance through its dual-circulation strategy, with a focus on resilient industrial and supply-chain capacities \cite{eu-open_strategic-autonomy}.

In this framework, startups, and particularly venture-backed startups, have become increasingly important because they are especially active in many of the emerging domains at stake, including AI, quantum computing, medtech, cybersecurity, cloud computing, and space technologies. While incumbents, dominant platforms, and states shape technological trajectories through scale, infrastructure, regulation, and procurement, venture-backed startups play a distinctive role in translating scientific advances into prototypes, products, and new industrial options. They therefore contribute to the development and potential control of technologies that matter for technological sovereignty.

However, despite this salience, systematic and comparable assessments of national performance in emerging technological sovereignty remain limited. Existing approaches are often qualitative (e.g., \cite{ecfr}) and lack a comparative quantitative framework that would allow countries and technologies to be benchmarked on consistent grounds. This constrains policymakers’ ability to monitor positioning, allocate resources, and design targeted interventions to strengthen strategic autonomy. These limitations motivate the need for a quantitative perspective that complements qualitative assessments and captures the structure of national positioning across emerging technological domains — an objective we pursue in the remainder of the paper.

Furthermore, Hawkins et al. \cite{hawkins2025_aiagency} have recently argued, in the case of AI, that the language of technological sovereignty can be conceptually confusing when compared with more operational policy questions about agency \cite{lehdonvirta2025_weaponized} and power in an interdependent world. In their framework, geoeconomic power is inherently \emph{relative}: strategic advantage depends on the structure of interdependence and on how one country’s position compares with that of others, rather than on the intrinsic importance of a technology considered in isolation. This perspective is closely aligned with the comparative portfolio-based approach developed in this paper across both countries and technologies. It insists on the fact that policymakers and analysts need to know not only whether a nation is present in a given emerging domain, but also how the \emph{directions} of its venture specializations compare with that of its peers, and which ones contribute most to its relative geoeconomic position. Technologies differ in their geoeconomic leverage: some are broadly developed and therefore weakly differentiating, whereas others remain concentrated in a smaller set of countries and thus carry greater geoeconomic weight. A systematic assessment of this relative positioning is therefore needed to move beyond narrative claims and to inform trade-offs, sequencing, and coordination in sovereignty strategies. In what follows, we address this need by constructing comparative, investment-based indicators of the geoeconomic potential of both countries and technologies.

\section{Quantifying relative geoeconomic power with an economic complexity approach}
\label{complexity}

The economic complexity framework, first introduced by Hidalgo and Hausmann \cite{ecomplx}, quantifies patterns of development through the diversity and sophistication of a nation’s productive capabilities. The central premise is that development depends not only on what an economy produces, but also on the structure of capabilities that enables production. Countries are thus not simple portfolios of sectors; they are embedded in networks of interdependent capabilities in which relative patterns of specialization matter \cite{product_space}. A core object is the country--activity \emph{specialization matrix}, which records whether a country is relatively specialized in a given activity \cite{atlas_ec, ecomplx, ecomplex}. Given this matrix, the complexity of both countries and activities can be inferred from the structure of their interactions.


Because of its generality, this framework has been adapted beyond trade-based applications, including analyses of regional disparities \cite{regiocomplex}, international innovation patterns \cite{lpdd, inoua, optimizing_eci}, and national ``software complexity'' \cite{softcomplex}. Closely related in spirit, but focused on inventive activity and R\&D, \cite{girolamo} apply economic-complexity ideas to ``complex technologies'' using patent-based indicators.

To assess the geoeconomic power of countries with respect to investments in the development and deployment of emerging technologies, we apply here this approach to venture capital (VC) portfolios, which track resource allocations to specific technological directions. Specifically, we analyze country--technology specializations derived from the relative allocation of VC to startups across emerging technological domains (Section~\ref{methods}). Formally, let \(M\) denote the country--technology specialization matrix, with \(M_{ct}=1\) if country \(c\) is specialized in technology \(t\) (e.g., according to an RCA threshold) and \(0\) otherwise. The standard economic complexity paradigm posits mutually reinforcing definitions: a technology is complex if it is specialized in by complex countries, and a country is complex if it specializes in complex technologies \cite{paradigm_eci, inoua, interpretation}. Let \(\mathbf{x}\) denote a vector of country complexities and \(\mathbf{y}\) a vector of technology complexities. Then
\begin{equation}
\begin{aligned}
\mathbf{x} &= \alpha\, M\, \mathbf{y},\\
\mathbf{y} &= \beta\, M^{\top}\, \mathbf{x},
\end{aligned}
\end{equation}
with \(\alpha,\beta>0\) scaling constants. Eliminating \(\mathbf{x}\) and \(\mathbf{y}\) yields the eigenvalue problems
\begin{equation}
\begin{aligned}
M M^{\top}\, \mathbf{x} &= (\alpha\beta)^{-1}\, \mathbf{x},\\
M^{\top} M\, \mathbf{y} &= (\alpha\beta)^{-1}\, \mathbf{y}.
\end{aligned}
\end{equation}
Here, \(M M^{\top}\) is the country--country matrix whose entries count co-specializations across technologies, and \(M^{\top} M\) is the technology--technology matrix whose entries count co-specializations across countries. The associated eigenvectors capture structure in the underlying bipartite network \cite{inoua, interpretation}. Following the canonical construction, we remove the leading (Perron--Frobenius) component, which primarily reflects overall diversification/ubiquity, and use the subsequent eigenvector(s) to define complexity measures for countries and technologies \cite{ecomplx, genepy}.

We derive two indices aligned with the standard framework. The \emph{Geoeconomic Complexity Index} (GCI) mirrors the Economic Complexity Index and ranks countries by the structure of their venture specializations across emerging technologies: high-GCI countries specialize in technologies that are geoeconomically complex and relatively rare. The \emph{Emerging Technology Geoeconomic Complexity Index} (ETGCI) mirrors the Product Complexity Index and ranks technologies by the extent to which their specialization is concentrated among high-GCI countries \cite{paradigm_eci, atlas_ec, interpretation}. Taken together, GCI and ETGCI provide a quantitative basis for evaluating the relative geoeconomic positioning of countries via their venture investment specializations, within the broader landscape of emerging technological sovereignty.

Geoeconomic complexity is therefore an \emph{emergent} property of the structure of the country--technology bipartite network. A technology attains high geoeconomic complexity if it is relatively uncommon (low ubiquity) and disproportionately selected by countries whose overall portfolios are themselves complex. Conversely, a technology that is widely pursued and adopted by countries with less complex portfolios will tend to exhibit lower geoeconomic complexity, regardless of its scientific or engineering difficulty. For instance, an emerging domain such as Cloud Computing can rank as highly complex in the geoeconomic sense (Section~\ref{results}) if only a small set of high-complexity countries specialize in it. By contrast, a domain such as Quantum Computing may appear less complex geoeconomically if it is a common specialization pursued by a broad set of countries, including many with less complex portfolios. Our analysis is therefore purely \emph{geoeconomic}: it does not address the other crucial but different and complementary issue about whether controlling a technology may be required for reasons that are specific to this particular technology, such as security in the case of Quantum Computing.

\section{Materials and Methods}
\label{methods}

\textit{Dataset.} We use Crunchbase — a widely adopted firm- and deal-level database that provides startup descriptions, headquarters locations, and round-level funding histories — as the primary source to build country–technology observations \cite{crunchbase}. To ensure consistency and adequate coverage, we restrict the sample to the 17 countries with the highest aggregate venture funding. For each country and year \(t\) from 2014 onward, we compile the list (up to 3{,}000 firms) of startups that raised the most capital, subject to a minimum of \(\$1\,\mathrm{M}\). Table~\ref{funding_table} reports, for each country, the number of startups thus retrieved through 2024 and the corresponding number of financing rounds. Fewer than 3{,}000 startups are selected for countries with fewer firms exceeding the \(\$1\,\mathrm{M}\) threshold.

\textit{Taxonomy.} Drawing on reports that identify emerging sovereign technologies (e.g., \cite{EIC_2023, EIC_2024, uk_gov_taxo}), we compile a list of 18 emerging technological domains (Section~\ref{list_sectors}). We then implement a probabilistic, multi-label classification procedure based on a large language model (ChatGPT-4o-mini) to assign each startup, using its Crunchbase description, to at most two domains from this list. Assignment probabilities are calibrated against a manually annotated validation set of 200 startups (at least 10 per domain), yielding mean precision of 90.9\% and mean recall of 78.8\%. Low-confidence assignments are excluded using an empirically selected probability threshold. Countries with fewer than 500 startups associated with at least one emerging domain in a given year are excluded from the analysis for that year.

\textit{Specialization matrix.} For each country \(i\), domain \(j\), and year \(t\), let \(S_{ij}^{t}\) denote the total venture capital invested during year \(t\) in startups headquartered in country \(i\) and associated with emerging technological domain \(j\). The Revealed Venture Advantage (RVA) that compares country \(i\)'s share of global investment in domain \(j\) to country \(i\)'s overall share of global venture investment is defined here in an identical way to standard RCA (Revealed Comparative Advantage) metrics for exports \cite{ecomplex} and to RTA (Revealed Technological Advantage) for patents \cite{soete}:
\begin{equation}
\label{eq:1}
M_{ij}^{t} =
\begin{cases}
1, & \displaystyle \frac{S_{ij}^{t}}{\sum_{i} S_{ij}^{t}} \;\ge\; \frac{\sum_{j} S_{ij}^{t}}{\sum_{i,j} S_{ij}^{t}},\\[0.8em]
0, & \text{otherwise.}
\end{cases}
\end{equation}

I.e., country \(i\) is specialized in technological domain \(j\) in year \(t\) if and only if \(M_{ij}^{t}=1\). RVAs and the associated country–technology matrix \(M^{t}\) therefore capture the favored technological directions of venture-backed innovation for countries in our dataset.

\textit{Threshold sensitivity and robustness.} As noted in prior work \cite{ecomplex, rethinking}, binarization can introduce threshold sensitivity. We implement two complementary approaches. (i) We numerically spread investment distributions by rounding each \(S_{ij}^{t}\) upward to the next \(\$100\,\mathrm{M}\) before computing \eqref{eq:1}; results using this approach are reported in Section~\ref{results} and Appendix~\ref{first_methodo}. (ii) We apply a two-year window (current year \(t\) and \(t-1\)) when determining specialization, following \cite{atlas_ec, ecomplex}; results are provided in Appendix~\ref{second_methodo} and are qualitatively similar. We also note that continuous, non-binarized formulations—such as ECI+ \cite{eci+} and GENEPY \cite{genepy}—have been proposed. Unreported analyses using GENEPY yield results consistent with the two approaches presented here.

\textit{Metrics and indices.} From the specialization matrix, we compute the following quantities \cite{paradigm_eci, atlas_ec, interpretation}:
\begin{itemize}
    \item \emph{Diversity} (country level): number of emerging technological domains in which a country is specialized.
    \item \emph{Ubiquity} (emerging technological domain level): number of countries specialized in a given emerging technological domain.
    \item \emph{Emerging Technology Geoeconomic Complexity Index (ETGCI)}: as defined in Section~\ref{complexity}, an emerging technological domain-level measure that captures its geoeconomic complexity.
    \item \emph{Geoeconomic Complexity Index (GCI)}: as defined in Section~\ref{complexity}, a country-level measure that captures the geoeconomic complexity of the portfolio of venture-backed technological directions selected by a given country.
\end{itemize}

\textit{Simulations and relatedness.} To identify potential strategic moves, we simulate every possible single new specialization for each country by toggling \(M_{ij}^{t}\) from 0 to 1 for domains \(j\) not currently specialized, recomputing the GCI and the implied rank change. Feasibility is proxied by technological relatedness in the sense of economic complexity \cite{paradigm_eci, product_space}: the proximity between a country and a domain increases with the frequency with which that domain co-occurs as a specialization with the existing specialization of the country in other countries. Combining impact (simulated GCI gain) and feasibility (relatedness) yields, for each country, the \emph{Simplest Single Sovereignty Enhancing Technology} (SSSET): the new specialization with the largest expected improvement in GCI rank among those with the highest relatedness.

\section{Results}
\label{results}

\begin{figure}[t]
  \centering
  \includegraphics[width=\columnwidth]{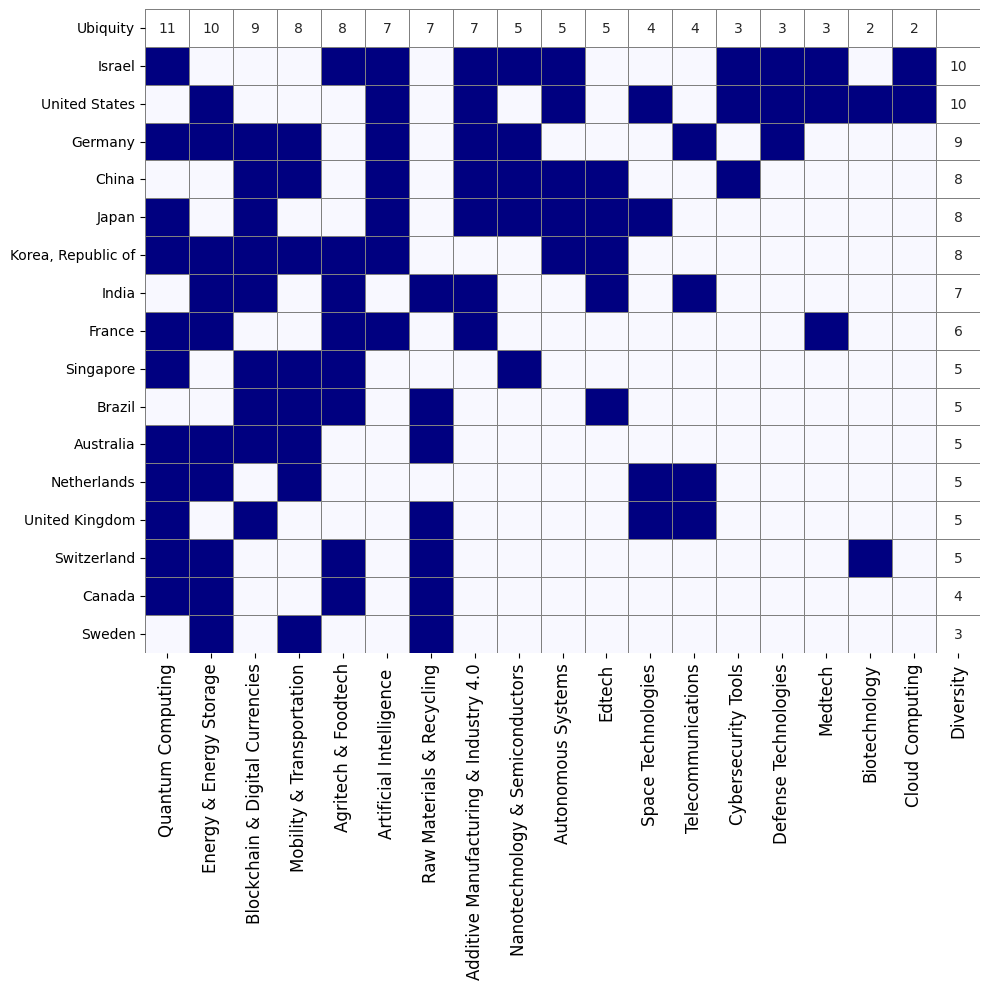}
  \caption{Country--domain specialization matrix in 2024. Rows (countries) are ordered by diversity; columns (emerging technological domains) by ubiquity.}
  \label{mcp_grid_heatmap}
\end{figure}

We first present the specialization matrix \(M\) and RVAs (Revealed Venture Advantage) for the year 2024 (Fig.~\ref{mcp_grid_heatmap}). Countries are ordered from top to bottom by decreasing \emph{diversity} (the number of their domains of specialization), while emerging technological domains are presented from left to right by decreasing \emph{ubiquity} (the number of countries specialized in each technological domain).

\begin{figure*}[t]
  \centering
  \includegraphics[width=\textwidth]{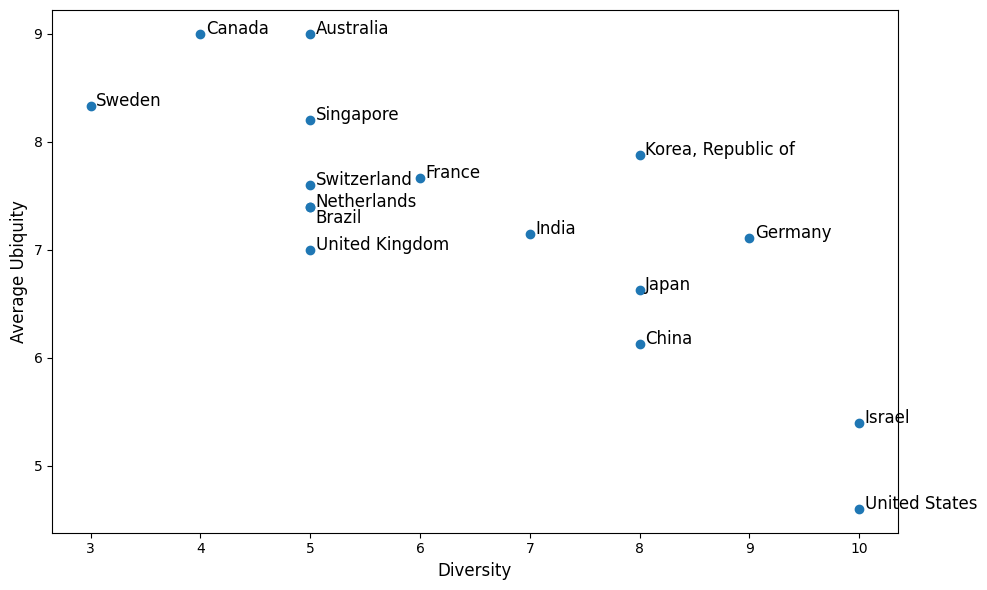}
  \caption{Country diversity (x-axis) versus the average ubiquity of its specialized domains (y-axis), 2024. Lower values on the y-axis indicate concentration of specializations in less ubiquitous technological domains.}
  \label{div_ubuqituy}
\end{figure*}

We then present this information at the country level by plotting each country’s diversity against the average ubiquity of the domains in which it is specialized (Fig.~\ref{div_ubuqituy}). According to this RVA perspective, Israel and the United States unambiguously stand out by combining the highest levels of diversity with the lowest average ubiquities. This combination places them in the extreme “high-diversity / low-ubiquity” region of the distribution, which indicates not only broad specialization but, moreover, a particularly strategic concentration of venture investments in less frequent technological domains. This strategic approach sharply differentiates their venture portfolios from those of the other countries in the sample: by contrast, Sweden exhibits the lowest diversity together with relatively high average ubiquity, consistent with specialization in domains that are more widely pursued, while countries such as France and the United Kingdom are associated with an average positioning that might result from a less proactive approach to the strategic element in international venture capital allocations. Germany, for its part, has a large diversity -- numerous specializations, according to RVAs -- but an average ubiquity of its specializations, while China is closest to Israel and the United States in its own positioning.

Table~\ref{sector_complexity} reports the Emerging Technology Geoeconomic Complexity Index (ETGCI), normalized to \([0,1]\). Cloud Computing, Cybersecurity Tools, and Medtech rank highest, reflecting that specialization in these domains is concentrated among a small set of high-GCI countries (Israel and the United States for Cloud Computing; Israel, the United States, and China for Cybersecurity Tools; Israel, the United States, and France for Medtech). By contrast, Artificial Intelligence, Quantum Computing, and Energy \& Energy Storage rank lower. This should not be interpreted as a statement about their intrinsic strategic importance. Rather, these domains are highly ubiquitous: many countries specialize in them, including several mid- or low-GCI ones, which mechanically lowers their ETGCI and thus their differentiating geoeconomic contribution to technological sovereignty. Put differently, there may be \emph{more} countries that can plausibly emerge as winners in these domains than in Cloud Computing or Cybersecurity Tools. At the same time, this does not preclude certain technologies from being intrinsically essential for sovereignty---whether through direct control enabled by defensive policies or through indirect dependence created by the breadth and specificity of their applications.

\begin{table}[t]
\centering
\footnotesize
\resizebox{\columnwidth}{!}{%
\begin{tabular}{lrr}
\hline
\textbf{Domains} & \textbf{Ubiquity} & \textbf{ETGCI} \\ \hline
Cloud Computing & 2 & 1.000 \\
Cybersecurity Tools & 3 & 0.895 \\
Medtech & 3 & 0.870 \\
Defense Technologies & 3 & 0.818 \\
Autonomous Systems & 5 & 0.723 \\
Artificial Intelligence (AI) & 7 & 0.668 \\
Additive Manufacturing \& Industry 4.0 & 7 & 0.637 \\
Nanotechnology \& Semiconductors & 5 & 0.590 \\
Biotechnology & 2 & 0.566 \\
Space Technologies & 4 & 0.426 \\
Edtech & 5 & 0.343 \\
Quantum Computing & 11 & 0.307 \\
Agritech \& Foodtech & 8 & 0.297 \\
Blockchain \& Digital Currencies & 9 & 0.263 \\
Energy \& Energy Storage & 10 & 0.250 \\
Mobility \& Transportation & 8 & 0.196 \\
Telecommunications & 4 & 0.171 \\
Raw Materials \& Recycling & 7 & 0.000 \\ \hline
\end{tabular}%
}
\caption{Ubiquity and ETGCI of emerging technological domains (2024).}
\label{sector_complexity}
\end{table}

Table~\ref{country_complexity} reports the Geoeconomic Complexity Index (GCI) by country, computed from the (normalized) ETGCI. The United States and Israel lead the ranking, followed by China, France, Japan, Germany, and the Republic of Korea. These countries combine comparatively broad specialization portfolios (e.g., 6 domains for France; 10 for Israel and the United States) with a marked emphasis on less ubiquitous, high-ETGCI emerging domains (e.g., Defense Technologies, Cloud Computing, Medtech, Cybersecurity Tools). Sweden and Australia lie at the lower end of the ranking, reflecting portfolios that are narrower and concentrated in more ubiquitous, lower-ETGCI domains (e.g., Energy \& Energy Storage, Mobility \& Transportation, Raw Materials \& Recycling). The United Kingdom ranks lower than several peer economies, consistent with a specialization profile tilted toward lower- and mid-ETGCI domains.

\begin{table}[t]
\centering
\begin{tabular}{lrr}
\hline
\textbf{Country} & \textbf{Diversity} & \textbf{GCI} \\ \hline
United States & 10 & 0.685 \\
Israel & 10 & 0.680 \\
China & 8 & 0.539 \\
France & 6 & 0.505 \\
Japan & 8 & 0.495 \\
Germany & 9 & 0.433 \\
Republic of Korea & 8 & 0.381 \\
Singapore & 5 & 0.331 \\
Switzerland & 5 & 0.284 \\
India & 7 & 0.280 \\
Netherlands & 5 & 0.270 \\
United Kingdom & 5 & 0.233 \\
Brazil & 5 & 0.220 \\
Canada & 4 & 0.214 \\
Australia & 5 & 0.203 \\
Sweden & 3 & 0.149 \\ \hline
\end{tabular}%
\caption{Country diversity and GCI (2024).}
\label{country_complexity}
\end{table}

Temporal patterns are comparatively stable from 2021 onward (Appendix~\ref{first_methodo}). Cloud Computing, Defense Technologies, and Cybersecurity Tools remain among the highest-ranked emerging technological domains, with a modest decline in Defense Technologies after 2022 and a concurrent rise in Medtech. Country rankings display similar persistence: the United States and Israel remain at the top throughout. France and Japan rank consistently high, reflecting sustained specializations in Medtech and AI (France) and in Autonomous Systems and AI (Japan). China maintains specializations in AI and Autonomous Systems (all years except 2021), and in Cybersecurity Tools (2020 and 2023--2024). By contrast, the United Kingdom’s ranking declines over time, consistent with the erosion of several earlier specializations.

Finally, we simulate single-addition specializations and evaluate their feasibility via relatedness (Section~\ref{methods}). Table~\ref{gci_best_close_prediction} reports the \emph{Simplest Single Sovereignty Enhancing Technology} (SSSET) for each country. While adding a specialization mechanically increases a country’s GCI score, the effects on \emph{rank} are limited at the top of the distribution: no single new specialization improves the relative ranking of the United States or Israel, nor of China and Germany. By contrast, several mid-ranked countries exhibit SSSETs with high relatedness and non-negligible rank gains. For example, Autonomous Systems (France) and AI (Singapore) yield the largest simulated improvement in rank (one position in both cases) while also being among the most related domains given their current specialization portfolios, suggesting comparatively accessible pathways for incremental upgrading. It should be noted, however, that relatedness is only one dimension of policy feasibility and desirability: countries may prioritize technologies for strategic or security reasons even when these are not the most related additions, particularly when control is deemed critical because of the specificity or breadth of downstream applications.

\begin{table*}[t]
\centering
\footnotesize
\resizebox{\textwidth}{!}{%
\begin{tabular}{llcc}
\hline
\textbf{Country} & \textbf{Simplest Single Sovereignty Enhancing Technology (SSSET)} & \textbf{Rank change} & \textbf{SSSET relatedness} \\ \hline
United States        & None                                                & =   & -- \\
Israel               & None                                                & =   & -- \\
China                & None                                                & =   & -- \\
France               & Autonomous Systems                                  & +1  & 2  \\
Japan                & Cybersecurity Tools                                 & +2  & 4  \\
Germany              & None                                                & =   & -- \\
Republic of Korea    & Defense Technologies                                & +1  & 6  \\
Singapore            & Artificial Intelligence                             & +1  & 2  \\
Switzerland          & Autonomous Systems                                  & +2  & 7  \\
India                & Artificial Intelligence                             & +2  & 3  \\
Netherlands          & Cybersecurity Tools                                 & +4  & 11 \\
United Kingdom       & Cybersecurity Tools or Medtech                & +4  & 10 \\
Brazil               & Artificial Intelligence or Additive Manufacturing \& Industry 4.0 & +4 & 3  \\
Canada               & Cybersecurity Tools                       & +6  & 13 \\
Australia            & Cybersecurity Tools or Medtech                      & +6  & 10 \\
Sweden               & Cybersecurity Tools or Medtech                      & +7  & 13 \\ \hline
\end{tabular}%
}
\caption{For each country, the SSSET (largest simulated GCI rank improvement for a single new specialization), the associated rank change, and its relatedness (feasibility proxy).}
\label{gci_best_close_prediction}
\end{table*}

\section{Limitations and future work}
\label{limitations}
In this study, causal claims are neither attempted nor implied; the results are descriptive and structural. Furthermore, venture portfolios capture one geoeconomic dimension of emerging technological sovereignty. Public R\&D, export capacity, standards leadership, supply-chain control, defense procurement, and talent stocks also bear on sovereignty but are not modeled here.

We would also like to stress several additional limitations. (1) \emph{Taxonomy and classification.} The 18-technological domain taxonomy and the LLM-assisted, multi-label assignment remain subject to misclassification and boundary shifts as technologies evolve. Further investigations should attempt to benefit from more advanced LLM models and to lower the granularity of the taxonomy by addressing technologies more specifically (e.g., nuclear technologies within Energy \& Energy Storage). (2) \emph{Coverage and measurement.} Crunchbase undercoverage and heterogeneity across countries may bias investment totals; results should be interpreted as lower-bound, venture-visible signals. (3) \emph{Binarization and thresholds.} Specialization relies on an RCA-style threshold with robustness checks (rounding and two-year windows). Continuous variants (e.g., ECI+, GENEPY) yield similar patterns in unreported tests, but sensitivity to thresholding remains a caveat. (4) \emph{Nationality attribution.} Most importantly, we assign nationality by startup headquarters to maintain comparability. This abstracts from investor nationality and control. Where foreign-controlled capital is significant, domestic sovereignty may be overstated; conversely, domestic control over foreign-headquartered firms is understated. Extending the framework to incorporate investor ownership and control is a priority for future work. 

Finally, special consideration should be given to blocs of countries, notably the European Union. In a bloc, member states may specialize in complementary emerging technological domains, so that bloc strength lies in portfolio coordination rather than in within-country breadth. As a result, a simple aggregation of venture flows at the bloc level may be misleading---especially in the present framework---because summing across countries dilutes specialization shares and can attenuate revealed comparative advantage. For instance, if country \(A\) is strongly specialized in technology \(X\) while country \(B\) invests only marginally in \(X\), pooling their flows mechanically reduces the bloc-level intensity in \(X\) and may erase \(A\)’s specialization signal in the bloc-level matrix (cf.\ Eq.~\ref{eq:1}). A more promising route would be to define bloc-level specialization using rule-based assignments---e.g., declaring a bloc specialized if at least one member, or a sufficiently large (or economically large) subset of members, is specialized. As a first illustration, and when applied to the France--Germany pair, even the most inclusive version of this rule (``any member specialized'') does not improve their combined ranking relative to France alone. This result is driven primarily by the absence of specializations in the highest-ETGCI domains (notably Cloud Computing and Cybersecurity Tools) in both countries, which might suggest that an EU27 technological sovereignty strategy may depend less on aggregation and more on coordinated complementarity, supported by shared instruments and joint efforts to build specializations in selected high-ETGCI domains.

\section{Conclusion}
\label{conclusion}

This article attempts to shed new light on emerging technological sovereignty by quantifying one of its key dimensions: the relative geoeconomic positioning associated with national venture portfolios in emerging technologies. It develops a quantitative framework that maps country–technology venture specializations into an economic-complexity setting, yielding two indices: the Geoeconomic Complexity Index (GCI) that ranks countries, and the Emerging Technology Geoeconomic Complexity Index (ETGCI) that ranks emerging technological domains. Their rankings are stable for recent years (2021-2024): for countries, the United States and Israel lead, followed by China, France, Japan and Germany, while Cloud Computing, Cybersecurity Tools, and Medtech are the most geoeconomically complex technological directions. Artificial Intelligence, Quantum Computing and Energy \& Energy Storage paradoxically rank below in this respect, highlighting the difference between the intrinsic importance of some technologies for sovereignty and their geoeconomic complexity potential, which is dependent upon the relative specializations adopted by national venture portfolios. Most notably, the geoeconomic potential of an emerging technological domain is enhanced when only a few, especially geoeconomically powerful players are involved, as in arms races, which seems to be especially the case for Cloud Computing, Cybersecurity Tools, and Medtech.

\onecolumngrid
\newpage
\section{Appendix}


\vspace{-2mm}
\subsection{List of emerging technological domains}\label{list_sectors}
\vspace{-2mm}

\begin{itemize}
\item \textbf{Domain 1, Artificial Intelligence (AI):} Natural language processing models - AI chips (e.g., neural processors) - AI frameworks for defense, healthcare, and national security - Ethical AI and Algorithmic Governance,
\item \textbf{Domain 2, Nanotechnology \& Semiconductors:} Chip design and manufacturing processes (e.g., photolithography) - Silicon fabrication facilities (fabs) - Foundry equipment (e.g. lithography machines) - Smart Factories - Microelectronics and Photonics,
\item \textbf{Domain 3, Cybersecurity Tools:} Encryption technologies - Secure communication platforms - Threat detection and response systems - Firewalls and cybersecurity frameworks,
\item \textbf{Domain 4, Space Technologies:} Satellite systems - Rocket propulsion systems - Space monitoring and defense platforms,
\item \textbf{Domain 5, Cloud Computing:} Cloud infrastructure - Data storage and processing platforms compliant with local regulations - Distributed computing systems for critical data,
\item \textbf{Domain 6, Telecommunications:} 5G/6G networks (e.g., Huawei alternatives for infrastructure) - Mobile Networks (or Wireless Networks) - Secure communication networks - Indigenous software-defined networking (SDN),
\item \textbf{Domain 7, Quantum Computing:} Quantum key distribution (QKD) for secure communication - National initiatives for quantum computer development - Quantum encryption standards,
\item \textbf{Domain 8, Energy \& Energy Storage:} Energy storage infrastructure - Advanced battery technologies - Smart Energy and Grid Management - Smart grid management systems - Solar, wind, and nuclear energy technologies - Hydropower and Hydrogen energy technologies - Fusion Energy,
\item \textbf{Domain 9, Autonomous Systems:} Autonomous vehicles (ground, air, and sea) - Non-Military Drones - Robotics for manufacturing and healthcare,
\item \textbf{Domain 10, Biotechnology:} Synthetic Biology - Regenerative Medicine - Microbial Biotechnology - Vaccine manufacturing capabilities,
\item \textbf{Domain 11, Defense Technologies:} Indigenous weapon systems (e.g., hypersonic missiles, stealth aircraft) - Command and control systems - Cyber-defense and intelligence platforms - Intelligence Analysis,
\item \textbf{Domain 12, Blockchain \& Digital Currencies:} Cryptocurrency and Decentralized Finance - Digital Art and NFT Platforms - Central Bank Digital Currencies (CBDCs),
\item \textbf{Domain 13, Additive Manufacturing (3D Printing) \& Industry 4.0:} 3D printing for aerospace, medical, and construction needs - Localized manufacturing systems to reduce dependence on imports,
\item \textbf{Domain 14, Mobility \& Transportation:} Electric vehicles - Hybrid vehicles - Electric Vehicle (EV) Infrastructure,
\item \textbf{Domain 15, Raw Materials \& Recycling:} Mining and Extraction - Advanced Biodegradable Plastics and Materials,
\item \textbf{Domain 16, Agritech \& Foodtech:} Advanced Urban Farming and Vertical Agriculture - Alternative Proteins - Food Waste Management,
\item \textbf{Domain 17, Medtech:} Diagnostic imaging (MRI, CT scans, ultrasound) - Prosthetics \& bionics - Surgical robotics \& AI-assisted surgery,
\item \textbf{Domain 18, Edtech:} Online Learning \& E-Learning Platforms - Virtual \& Augmented Reality (VR/AR) in Education.
\end{itemize}

\vspace{-2mm}

\subsection{Number of startups, of funding rounds and total raised per country (as of 2024)}\label{funding_table}

\begin{table}[H]
\begin{spacing}{1.5}

\resizebox{\textwidth}{!}{%

\begin{tabular}{l*{18}{c}c}
 & \textbf{Number of startups} & \textbf{Number of funding rounds} & \textbf{Total raised (B\$)} \\
\hline
\textbf{Australia} & 1,348 & 3,464 & 75.8 \\
\textbf{Brazil} & 646 & 1,836 & 46.3 \\
\textbf{Canada} & 3,073 & 11,605 & 184.9 \\
\textbf{China} & 3,712 & 10,974 & 572.5 \\
\textbf{France} & 2,582 & 6,486 & 89.9 \\
\textbf{Germany} & 2,109 & 6,369 & 109.5 \\
\textbf{India} & 2,121 & 6,849 & 227.1 \\
\textbf{Israel} & 1,834 & 5,468 & 62.2 \\
\textbf{Japan} & 1,230 & 4,088 & 40 \\
\textbf{Republic of Korea} & 1,313 & 3,768 & 54.4 \\
\textbf{Netherlands} & 987 & 2,633 & 82.3 \\
\textbf{Singapore} & 1,233 & 3,284 & 68.4 \\
\textbf{Sweden} & 972 & 2,966 & 55.4 \\
\textbf{Switzerland} & 1,120 & 4,247 & 73.2 \\
\textbf{United Kingdom} & 3,555 & 12,817 & 269.3 \\
\textbf{United States} & 4,819 & 29,137 & 2,294.4 \\
\hline
\textbf{Total} & \textbf{32,654} & \textbf{115,991} & \textbf{4,305.4}\\

\end{tabular}%
}
\end{spacing}
\label{tab:country_Domain_contributions}
\end{table}

\newpage
\subsection{GCI \& ETGCI rankings, 2020 to 2024 (investments per country and domain rounded to the upper hundred million)} \label{first_methodo}

\twocolumngrid

\vspace*{\fill}

\begin{table}[H]
\centering

\rotatebox{90}{

\resizebox{\textwidth}{!}{
\begin{tabular}{ccccc}
\hline
\textbf{2020} & \textbf{2021} & \textbf{2022} & \textbf{2023} & \textbf{2024} \\ \hline

China & United States & United States & United States & United States \\ 
United States & Israel & Israel & Israel & Israel \\ 
United Kingdom & Japan & Japan & Japan & China \\ 
Israel & United Kingdom & Switzerland & France & France \\ 
Republic of Korea & France & China & Switzerland & Japan \\ 
Japan & China & France & China & Germany \\ 
Singapore & Republic of Korea & Germany & Sweden & Republic of Korea \\ 
France & Switzerland & United Kingdom & Germany & Singapore \\ 
India & Germany & Netherlands & Canada & Switzerland \\ 
Germany & Netherlands & Republic of Korea & Republic of Korea & India \\ 
Sweden & Sweden & India & Netherlands & Netherlands \\ 
Switzerland & India & Canada & United Kingdom & United Kingdom \\ 
Australia & Canada & Singapore & Australia & Brazil \\ 
Canada & Singapore & Australia & Singapore & Canada \\ 
 & Australia & Brazil & India & Australia \\ 
 &  & Sweden & Brazil & Sweden \\
\hline

\end{tabular}
}
}
\end{table}

\vfill

\begin{table}[H]

\centering
\rotatebox{90}{
\resizebox{1.4\textwidth}{!}{
\begin{tabular}{ccccc}
\hline
\textbf{2020} & \textbf{2021} & \textbf{2022} & \textbf{2023} & \textbf{2024} \\ \hline
Cybersecurity Tools & Cloud Computing & Defense Technologies & Cloud Computing & Cloud Computing \\
Cloud Computing & Defense Technologies & Cloud Computing & Biotechnology & Cybersecurity Tools \\
Autonomous Systems & Autonomous Systems & Cybersecurity Tools & Defense Technologies & Medtech \\
Space Technologies & Cybersecurity Tools & Medtech & Medtech & Defense Technologies \\
Telecommunications & Space Technologies & Autonomous Systems & Autonomous Systems & Autonomous Systems \\
Artificial Intelligence (AI) & Medtech & Biotechnology & Cybersecurity Tools & Artificial Intelligence (AI) \\
Edtech & Biotechnology & Space Technologies & Artificial Intelligence (AI) & Additive Manufacturing \& Industry 4.0 \\
Medtech & Additive Manufacturing \& Industry 4.0 & Additive Manufacturing \& Industry 4.0 & Space Technologies & Nanotechnology \& Semiconductors \\
Mobility \& Transportation & Artificial Intelligence (AI) & Nanotechnology \& Semiconductors & Nanotechnology \& Semiconductors & Biotechnology \\
Defense Technologies & Nanotechnology \& Semiconductors & Artificial Intelligence (AI) & Additive Manufacturing \& Industry 4.0 & Space Technologies \\
Nanotechnology \& Semiconductors & Agritech \& Foodtech & Quantum Computing & Energy \& Energy Storage & Edtech \\
Biotechnology & Edtech & Energy \& Energy Storage & Quantum Computing & Quantum Computing \\
Blockchain \& Digital Currencies & Telecommunications & Blockchain \& Digital Currencies & Agritech \& Foodtech & Agritech \& Foodtech \\
Agritech \& Foodtech & Quantum Computing & Agritech \& Foodtech & Raw Materials \& Recycling & Blockchain \& Digital Currencies \\
Additive Manufacturing \& Industry 4.0 & Blockchain \& Digital Currencies & Telecommunications & Mobility \& Transportation & Energy \& Energy Storage \\
Quantum Computing & Energy \& Energy Storage & Edtech & Blockchain \& Digital Currencies & Mobility \& Transportation \\
Energy \& Energy Storage & Raw Materials \& Recycling & Mobility \& Transportation & Edtech & Telecommunications \\
Raw Materials \& Recycling & Mobility \& Transportation & Raw Materials \& Recycling & Telecommunications & Raw Materials \& Recycling \\ \hline

\end{tabular}
}
}

\end{table}

\onecolumngrid

\newpage
\subsection{GCI \& ETGCI rankings, 2020 to 2024 (specialization considered over two consecutive year periods)} \label{second_methodo}

\twocolumngrid

\vspace*{\fill}

\begin{table}[H]

\centering
\rotatebox{90}{

\resizebox{\textwidth}{!}{

\begin{tabular}{ccccc}
\hline
\textbf{2020} & \textbf{2021} & \textbf{2022} & \textbf{2023} & \textbf{2024} \\ \hline
United States & United States & United States & United States & United States \\
Israel & Israel & Israel & Israel & Israel \\
United Kingdom & United Kingdom & Japan & Japan & China \\
Japan & Japan & United Kingdom & China & Japan \\
France & China & China & Switzerland & France \\
Switzerland & France & Switzerland & Republic of Korea & Germany \\
Sweden & Singapore & France & France & Republic of Korea\\
Australia & Germany & Netherlands & United Kingdom & Switzerland \\
Canada & Switzerland & Germany & Germany & Sweden \\
China & Sweden & Sweden & Netherlands & Singapore \\
Singapore & Republic of Korea & Republic of Korea & Sweden & Netherlands \\
Germany & Canada & Canada & India & India \\
Republic of Korea & Netherlands & Australia & Canada & Canada \\
India & India & India & Australia & Australia \\
 & Australia & Singapore & Singapore & United Kingdom \\
 &  & Brazil & Brazil & Brazil \\

\end{tabular}
}
}

\end{table}
\vfill

\begin{table}[H]
\centering

\rotatebox{90}{

\resizebox{1.4\textwidth}{!}{

\begin{tabular}{ccccc}
\hline
\textbf{2020} & \textbf{2021} & \textbf{2022} & \textbf{2023} & \textbf{2024} \\ \hline
Defense Technologies & Defense Technologies & Defense Technologies & Cloud Computing & Cloud Computing \\
Cloud Computing & Cloud Computing & Cloud Computing & Defense Technologies & Defense Technologies \\
Medtech & Autonomous Systems & Autonomous Systems & Autonomous Systems & Cybersecurity Tools \\
Autonomous Systems & Cybersecurity Tools & Cybersecurity Tools & Cybersecurity Tools & Autonomous Systems \\
Cybersecurity Tools & Medtech & Space Technologies & Space Technologies & Medtech \\
Space Technologies & Space Technologies & Medtech & Medtech & Biotechnology \\
Telecommunications & Telecommunications & Nanotechnology \& Semiconductors & Biotechnology & Artificial Intelligence (AI) \\
Quantum Computing & Quantum Computing & Artificial Intelligence (AI) & Additive Manufacturing \& Industry 4.0 & Additive Manufacturing \& Industry 4.0 \\
Nanotechnology \& Semiconductors & Artificial Intelligence (AI) & Quantum Computing & Artificial Intelligence (AI) & Nanotechnology \& Semiconductors \\
Additive Manufacturing \& Industry 4.0 & Additive Manufacturing \& Industry 4.0 & Additive Manufacturing \& Industry 4.0 & Nanotechnology \& Semiconductors & Space Technologies \\
Energy \& Energy Storage & Biotechnology & Biotechnology & Quantum Computing & Edtech \\
Biotechnology & Nanotechnology \& Semiconductors & Telecommunications & Blockchain \& Digital Currencies & Quantum Computing \\
Artificial Intelligence (AI) & Agritech \& Foodtech & Blockchain \& Digital Currencies & Energy \& Energy Storage & Agritech \& Foodtech \\
Raw Materials \& Recycling & Blockchain \& Digital Currencies & Raw Materials \& Recycling & Agritech \& Foodtech & Energy \& Energy Storage \\
Blockchain \& Digital Currencies & Energy \& Energy Storage & Energy \& Energy Storage & Edtech & Blockchain \& Digital Currencies \\
Agritech \& Foodtech & Mobility \& Transportation & Agritech \& Foodtech & Mobility \& Transportation & Mobility \& Transportation \\
Mobility \& Transportation & Edtech & Mobility \& Transportation & Raw Materials \& Recycling & Telecommunications \\
Edtech & Raw Materials \& Recycling & Edtech & Telecommunications & Raw Materials \& Recycling \\
\hline

\end{tabular}
}}
\end{table}

\onecolumngrid

\end{document}